\documentclass[twocolumn,showpacs,aps,pre,
groupedaddress,amssymb,amsmath,nobalancelastpage]{revtex4}

\usepackage{graphicx}% Include figure files
\usepackage{longtable}% Include figure files

\begin{document}

%\preprint{APS/123-QED}

\title{Statistics of Bubble Rearrangements in a Slowly Sheared Two-dimensional Foam}

\author{Michael Dennin}
\affiliation{Department of Physics and Astronomy, University of
California at Irvine, Irvine, California 92697-4575}

\date{\today}

\begin{abstract}

Many physical systems exhibit plastic flow when subjected to slow
steady shear. A unified picture of plastic flow is still lacking;
however, there is an emerging theoretical understanding of such
flows based on irreversible motions of the constituent
``particles'' of the material. Depending on the specific system,
various irreversible events have been studied, such as T1 events
in foam and shear transformation zones (STZ's) in amorphous
solids. This paper presents an experimental study of the T1 events
in a model, two-dimensional foam: bubble rafts. In particular, I
report on the connection between the distribution of T1 events and
the behavior of the average stress and average velocity profiles
during both the initial elastic response of the bubble raft and
the subsequent plastic flow at sufficiently high strains.

\end{abstract}

%\keywords{}

\maketitle

\section{Introduction}

Bubble rafts have been used as a model experimental system for the
study of crystalline and amorphous solids \cite{BL49,AK79} and for
the study of two-dimensional foam \cite{KE99,LTD02}. This overlap
is just one of many examples that points to an important question
in the study of the mechanical response of materials. Under
conditions of slow steady shear, what, if any, is the connection
between the response of ``mesoscopic'' materials, such as foams,
emulsions, pastes, and slurries, and plastic flow of ``molecular''
systems, such as amorphous solids? Based on macroscopic
measurements, the systems are similar. There is an initial elastic
response for small strains (or stresses) and a yield stress, above
which irreversible deformations, or plastic deformations, occur.
Eventually, above some critical stress (or strain), the system
enters a ``flowing'' state that is characterized by irregular
periods of stress increase and decrease. This is often referred to
as unbounded plastic flow. For the purposes of this paper, this
will simply be referred to as plastic flow. As one reduces the
shear rate, the critical stress approaches the yield stress in
such a way that for sufficiently slow shear rates the behavior of
the system is essentially shear-rate independent. This is often
referred to as the {\it quasistatic regime}. A complete
``microscopic'' picture of plastic flow still does not exist,
where microscopic refers to the fundamental length scale relevant
to the system in question. For example, in bubble rafts, it would
be the dynamics of individual bubbles. Open questions include: the
microscopic source of the stress release events; the spatial and
temporal distribution of such events; and the nature of such
events during periods of stress increase. Experimentally, the
challenge is identifying systems for which the microscopic events
are directly observable. This is one of the main advantages of
mesoscopic systems, such as the bubble raft, and the reason for
the interest in making connections between mesoscopic systems and
molecular systems, such as amorphous solids.

\begin{figure}
\includegraphics[width=8cm]{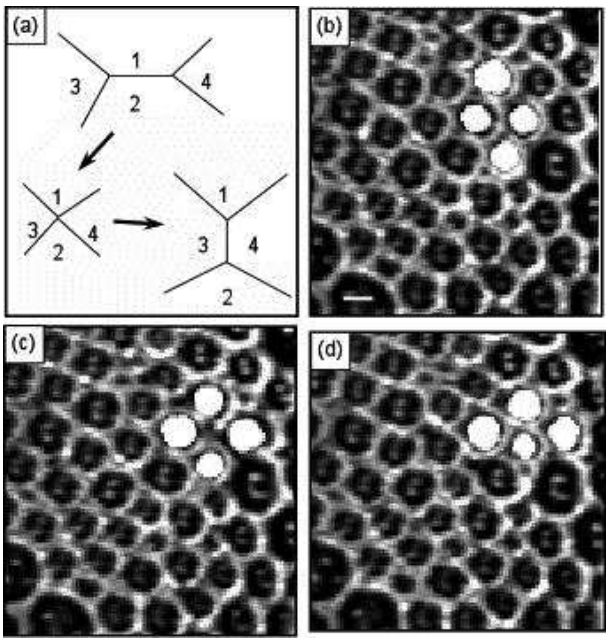}
\caption{\label{defT1} (a) Schematic representation of the three
main steps in a T1 event. The bubbles labelled 1 and 2 are
initially neighbors. As the bubbles are sheared, all four bubbles
meet at a vertex. After the event, the bubbles labelled 3 and 4
are neighbors. (b) - (d) are three images illustrating an actual
T1 event in the bubble raft. The location of one T1 event is
highlighted by artificially coloring the bubbles involved white.
The images are taken 3.2~s apart and the white scale bar in (b) is
2~mm long.}
\end{figure}

Models and simulations of diverse systems, ranging from solids
\cite{SVE83,BA94a,BA94b,BA94c,FL98,BVR02,O03,PALB04} to foam
\cite{KNN89,KOKN95,OK95,D95,D97,WBHA92,HWB95,JSSAG99,TSDKLL99,KD03},
have provided a number of insights into these questions. The focus
of this paper is on the role of T1 events in foam. Aqueous foam
consists of gas bubbles separated by liquid walls
\cite{S93,K88,WH99}. A T1 event is an irreversible neighbor
switching event. Both a schematic representation and an actual T1
event are presented in Fig.~\ref{defT1}. For the purposes of this
paper, we will focus on the role of T1 events during the steady
shear of foam. However, it should be mentioned that understanding
the nonlinear events that are {\it not} shear induced may be
important when comparing foam and amorphous solids. In the absence
of external stress, foam coarsens, and T1 events occur due to
geometric changes in the foam structure. These T1 events are not
necessarily distinguishable from those caused by flow. In
contrast, most amorphous solids do not exhibit the equivalent of
coarsening. However, thermally activated events may be important.
Possible differences between thermally activated and coarsening
events is an interesting question, but one beyond the scope of
this paper.

Simulations of flowing foam have characterized different aspects
of T1 events. Often, one separately considers T1 events and
``avalanches'', i.e. sudden releases of stress (or energy) in the
foam. One issue is whether or not the probability of the number of
T1 events in a given avalanche exhibits power-law behavior. A
common model to study dry foam (where the bubbles are essentially
polygonal) is the vertex model \cite{KNN89,KOKN95,OK95}.  In the
vertex model, a T1 event is defined to occur when the distance
between two vertices (i.e. the wall between two bubbles) is below
a threshold value. One then eliminates this wall, creating a T1
event. Therefore, within this model, all T1 events are essentially
instantaneous. In this case, simulations found evidence for
power-law behavior of the probability of T1 events. Another
characterization of the T1 events is the number of T1 events per
bubble per unit strain, $R_{T1}$. For the vertex model, this
quantity is $R_{T1} = 0.5$. A modified version of the vertex model
has been used to study the issue of flow localization under shear
\cite{KD03}. These simulations report a correlation between the
spatial localization of T1 events to the neighborhood of a system
boundary with the localization of shear in the same region. The
issue of shear localization will be discussed in more detail
later.

Another class of models focuses on wet foam (foam in which the
bubbles are essentially spherical, or in two-dimensions,
circular), using a quasistatic simulation \cite{WBHA92,HWB95}.
These simulations involve making a small step strain and then
allowing the system to relax to an energy minimum before applying
the next step strain. Anytime the energy decreases after a step
strain, one declares this an ``avalanche'' or ``event''. By
comparing neighbors in the initial and final state, one can count
the number of T1 events for a particular avalanche. In these
simulations, avalanches consisting of a large number of T1 events
were observed, suggesting the possibility of power-law behavior
\cite{WBHA92,HWB95}. For this model, $R_{T1}$ was not reported.

Wet foam under steady shear has also been simulated using a
q-Potts model \cite{JSSAG99}. In this case, different bubbles are
identified by different spin orientations. Simulations of the
q-Potts model find that the distribution of topological
rearrangements are not power-law-like. However, the distribution
of energy drops may be consistent with power-law behavior
\cite{JSSAG99}.

Another important set of simulations for wet foam involved
studying the steady shear of the bubble model
\cite{D95,D97,TSDKLL99}. This model treats bubbles as spheres (or
circles) that interact via a spring force proportional to their
overlap and a viscous drag proportional to velocity differences.
As this model directly simulates dynamics, the duration of T1
events exhibits a distribution of duration times. Simulations
focused on small shear-rates in the quasistatic limit, i.e. the
flow properties were independent of the shear rate. Under these
conditions, no evidence of power-law behavior is observed in the
bubble model at high bubble density, and $R_{T1} = 0.15$
\cite{D95,D97,TSDKLL99}. If one decreases the density of the
bubbles, it appears that the distribution of events approaches a
power-law as one approaches the critical density for the
``melting'' of the foam \cite{TSDKLL99}.

Before discussing the current state of experiments in foam, it is
useful to put the theoretical work on T1 events in foam in the
context of two points of view of plasticity in amorphous
materials. First, the idea of shear transformation zones (STZ's),
as developed by Falk and Langer \cite{FL98}, has received
significant attention. STZ are a way of describing local,
irreversible rearrangements of particles during shear.  STZ are
based on previous work by Spaepen and Argon on activated
transitions and Turnball, Cohen, and others on free-volume
fluctuations. As the STZ refers to a small region of the material
with certain properties \cite{A79}, there is only a loose
connection between the STZ and T1 events. However, it is
reasonable to identify as an STZ regions in which a few T1 events
combine to form a local slip (see, for instance,
Fig.~\ref{t1elastic}e). It is expected that the local
rearrangements identified as STZ are associated with quadrupolar
energy fluctuations. In fact, the expected quadrupolar energy
fluctuations have been observed associated with T1 events in a
simulation of foam \cite{KD03}, but not, as of yet, in simulations
of molecular systems.

Another view of plasticity is based on shear induced changes in
the potential energy landscape, as proposed by Malandro and Lacks
\cite{ML99}. This picture derives from an inherent structure
formalism and focuses on changes in the macroscopic mechanical
response of a material due to shear induced changes of the
potential energy. This formulism has been used to study
simulations of a quasistatic version of the bubble model
\cite{ML04}. In this case, system wide rearrangement events are
observed. This is not seen in bubble model simulations of the
quasistatic limit, but it is seen in other quasistatic simulations
of foam. The work in Ref.~\cite{ML04} suggests the need to
carefully define the concept of an ``event'', especially for
steady-state experiments where the time scale for events to occur
relative to the applied shear can be important. For example, a
shear-rate regime may exist that is quasistatic as defined by the
behavior of quantities such as the average stress, but not in a
quasistatic with regard to the duration of stress releases. Hence,
large events get ``broken up'' by the steady shear, changing the
nature of the distribution of events.

A number of experimental studies of bubble rearrangements in model
foam have been carried out. As mentioned, some of the earliest
work was done using bubble rafts \cite{BL49,AK79,MGC89}, i.e.
layers of gas bubbles floating on a liquid surface, as a model
molecular system, both for crystalline and amorphous solids
\cite{BL49,AK79}. One major advantage of bubble rafts is that
their two dimensional nature allows for easy imaging and tracking
of all of the ``particles'' in the system. Another reason that
bubble rafts have been so useful in the study of molecular systems
is that there exists detailed calculations of the bubble
interactions \cite{SA82}. More recently, bubble rafts were used to
study rearrangements after a step strain in order to make
comparisons with the quasistatic simulations of foam \cite{KE99}.
This work did not directly measure T1 events, but it did look at
changes in the number of neighbors for bubbles. The results
suggested that large scale events were possible.

Experiments have also been carried out using monolayer foam
\cite{DK97}. Langmuir monolayers consist of a single layer of
molecules confined to the air-water interface. They exhibit a
large number of two-dimensional phases, including gas-liquid
coexistence. This allows for the formation of a foam of gas
bubbles with liquid walls. For a monolayer foam under steady
shear, only a small number of simultaneous T1 events were
observed, with $R_{T1} = 0.12 \pm 0.03$. These results are
consistent with the bubble model.

As mentioned, the other aspect of T1 dynamics is their potential
role in explaining shear localization in yield-stress materials
\cite{KD03}, such as foam and granular systems. It has long been
known that a yield stress and/or nonlinear viscosity can lead to
inhomogeneous flows \cite{BOOKS}. However, it is only recently
that experimental techniques have allowed for detailed
measurements of such behavior. A number of such studies have been
carried out in granular materials, where exponential velocity
profiles (or other strongly localized velocity profiles) are
generally observed \cite{HBV99,LBLG00,MDKENJ00}. In contrast,
measurements in various three-dimensional pastes, slurries, and
foams show a different type of inhomogeneous flow. In this case,
the flow is not strongly localized, and there is a
shear-discontinuity at the boundary between flow and no flow
\cite{CRBMGH02,DCBC02}.

For two-dimensional foams, the situation is ambiguous.
Three-dimensional foam that is confined between plates to form a
model two-dimensional system exhibits shear localization analogous
to granular systems \cite{DTM01}. This work motivated simulations
of the modified vertex model discussed earlier that showed a
connection between T1 events and shear localization \cite{KD03}.
In this case, it appears that the spatial distribution of stress
released by the T1 event results in the subsequent localization of
the events. The localization of T1 events is correlated with the
localization of flow. In contrast, experiments with a bubble raft
exhibit a shear-discontinuity \cite{LCD04} similiar to that
reported in Refs.~\cite{CRBMGH02,DCBC02}. In Ref.~\cite{LCD04}, T1
events were not measured.

The work reported in this paper addresses the general question of
the temporal and spatial distribution of T1 events during the
slow, steady shear of a bubble raft. Also, connections between the
T1 events and the velocity profiles reported in Ref.~\cite{LCD04}
are made. The rest of the paper is organized as follows.
Section~\ref{ExpMeth} provides the details of the experimental
setup. The results are presented in two parts.
Section~\ref{elasticsec} presents the initial response of the
system. Section~\ref{plasticsec} presents the behavior during
plastic flow. Finally, the results are summarized and discussed in
Sec.~\ref{summary}.

\section{Experimental Methods}
\label{ExpMeth}

\begin{figure}
\includegraphics[width=8cm]{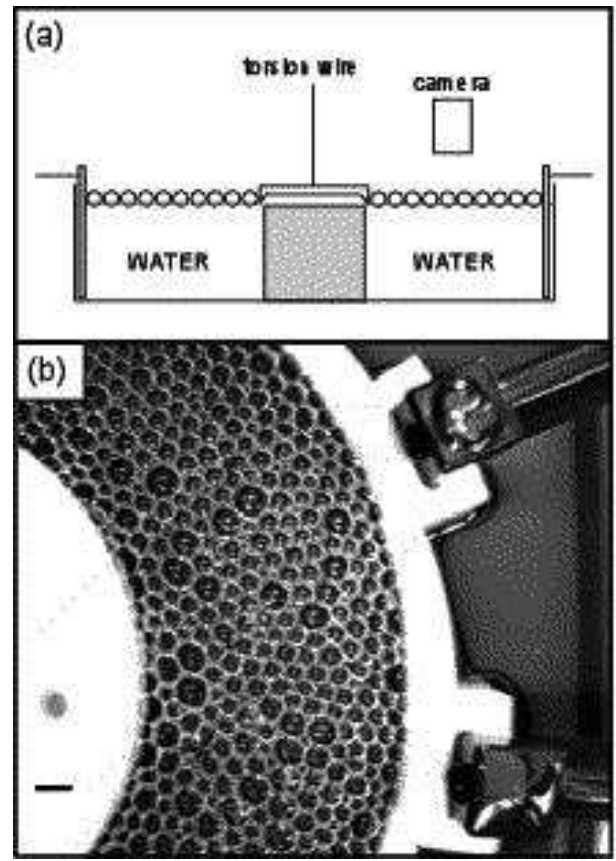}
\caption{\label{schematic} (a) A schematic drawing of the
apparatus showing a side view. The main elements of the apparatus
are the knife edge disk that is supported by a torsion wire and
that serves as the inner cylinder for the bubbles. There is a
separate fixed inner cylinder in the fluid (in gray). There is a
segmented outer cylinder for generating flow, and there is a fixed
dish that holds the fluid. The bubbles sit on top of the fluid, as
indicated by the circles. (b) An image from the top of the bubbles
in the apparatus that shows a portion of both the outer and inner
cylinder. The black scale bar in the lower left corner is 5~mm.}
\end{figure}

The experimental system consisted of a standard bubble raft
\cite{AK79} in a Couette geometry. The bubble raft was produced by
flowing regulated nitrogen gas through a hypodermic needle into a
homogeneous solution of 80\% by volume deionized water, 15\% by
volume glycerine, and 5.0\% by volume Miracle Bubbles (Imperial
Toy Corp.). The bubble size was dependent on the nitrogen flow
rate, which was varied using a needle valve.  A random
distribution of bubble sizes was used, with an average radius of
$1\ {\rm mm}$. The resulting bubbles were spooned into a
cylindrical Couette viscometer. This produced a two-dimensional
model of a wet foam on a homogeneous liquid substrate.
Figure~\ref{schematic} presents a schematic side view of the
bubbles in the apparatus and an image of a top view of the bubble
raft.

Due to the nature of the bubble raft, no measurable coarsening was
observed. However, after approximately two hours significant
numbers of bubbles would pop, presumably due to loss of fluid in
the walls. This set the upper limit on the total time of the
measurements. In contrast, during the initial two hour period,
only two out of approximately 400 bubbles in the field of view
were observed to pop.

An important feature of the bubble raft is the gas area fraction.
To achieve a desired gas area fraction, the bubble raft was
constructed by placing the approximate number of desired bubbles
in the trough with the outer barrier set to a large radius. It is
important to note that the bubbles exhibited a strong attraction
to each other. The outer barrier was compressed until the desired
radius was reached. The gas area fraction was determined by
thresholding images of the bubbles and counting the area inside of
the bubbles. Because of the three-dimensional nature of the
bubbles, this represents an operational definition of gas-area
fraction based on the details of the image analysis. However, the
choice of threshold was consistent with an estimate of the gas
area fraction based on the area of trough and expected
distribution of bubble sizes. For all of the data reported here,
the gas area fraction was approximately 0.95.

The Couette viscometer is described in detail in Ref.~\cite{app}
and shown schematically in Fig.~\ref{schematic}(a). It consists of
a shallow dish that contains the liquid substrate. Two concentric
Teflon barriers are placed vertically in the dish. Sections of
both of these barriers are visible in Fig.~\ref{schematic}(b). The
outer barrier is a ring consisting of twelve segmented pieces. It
has an adjustable radius. For the experiments discussed here, the
outer radius was fixed at $r_o = 7.43\ {\rm cm}$. The inner
barrier, or rotor, is a Teflon disk with a radius $r_i = 3.84\
{\rm cm}$. The outer edge of the disk is a knife edge that is just
in contact with the water surface. It was suspended by a wire to
form a torsion pendulum.

To shear the bubble raft, the outer Teflon barrier was rotated at
a constant angular velocity $\Omega = 8 \times 10^{-4}\ {\rm
rad/s}$. The first layer of bubbles at either boundary did not
slip relative to the boundary. Due to the finite size of the
bubbles, this results in an effective inner radius of $r = 4.4\
{\rm cm}$. Due to the cylindrical geometry, the shear rate is not
uniform across the system and is given by $\dot{\gamma}(r) = r
\frac{d}{dr}\frac{v(r)}{r}$. Here $v(r)$ is the azimuthal velocity
of the bubbles. During plastic flow, the average azimuthal
velocity of the inner cylinder is zero. Measurements of the
average azimuthal velocity profile allows for calculations of the
shear rate. As measured at $r = 4.4\ {\rm cm}$, $\dot{\gamma} = 4
\times 10^{-3}\ {\rm s^{-1}}$. In this regime, where reported, the
strain ($\gamma$) is taken to be the strain at this radius, and is
computed from $\gamma = \dot{\gamma}t + \gamma_o$, where $t$ is
the time since the initiation of plastic flow and $\gamma_o$ is
the amount of strain developed during the initial period. During
the initial period, the inner barrier has a finite angular speed.
However, one can still compute the effective shear rate at $r =
4.4\ {\rm cm}$. In this regime, $\dot{\gamma} = 3 \times 10^{-4}\
{\rm s^{-1}}$. Again, where reported, the strain is the strain at
the inner cylinder: $\gamma = \dot{\gamma}t$, where in this case
$t$ is measured from the initiation of shear.

The details of the velocity measurements are given in
Ref~\cite{LCD04}. Video tape of roughly one third of the trough
was recorded. Images from this tape were taken every 3.2 s and
digitized. An image processing routine based on standard
Labwindows functions was developed that detected and tracked
individual bubbles. This tracking software was also used to
compute the average bubble displacements. This is used to compute
the deviation of the bubble motion from ideal elastic behavior, as
discussed in Sec.~\ref{elasticsec}.

The T1 events were measured by stepping the digitized images one
frame at a time and visually searching for the location and time
at which T1 events occurred. A T1 event was defined to occur when
two bubbles were observed to lose contact, and two other bubbles
moved into the resulting space (see Fig.~\ref{defT1}). Due to the
associated motions of the other neighboring bubbles, T1 events are
relatively easy to detect by hand \cite{test}. For automatic
tracking of T1 events, it is critical to accurately detect
essentially all of the bubbles, as one is interested in
determining neighbor switching events. This is in contrast with
the displacement and velocity measurements where the requirement
is that one tracks enough bubbles to have sufficient statistics.
These are the reasons that automatic methods were used for
displacement and velocity measurements, but the detection of T1
events was by hand.

The stress on the inner rotor was determined using two different
methods. In both cases, the torque, $T = \kappa \theta$, on the
inner rotor is determined by measuring the angular displacement,
$\theta$. (For the experiments presented here, the torsion
constant $\kappa = 5.7 \times 10^{-7}\ {\rm N m}$.) The stress is
then determined from $\sigma = T/2\pi r^2$. The difference in the
two methods is the determination of $\theta$. The first method
uses a magnetic flux technique, and the details of this technique
are in Ref.~\cite{app}. This is the more precise of the two
methods, with a stress resolution of $3 \times 10^{-3}\ {\rm
mN/m}$. The second method uses the video images of the inner
cylinder and tracks fixed features on the disk. This method has a
resolution of $0.043\ {\rm mN/m}$. The second method is used to
correlate the video analysis of bubble motions (displacements and
T1 events) with the detailed stress fluctuations determined from
the magnetic flux measurements that are reported in
Ref.~\cite{PD03}.

As mentioned, foams are inherently nonequilibrium systems. One
complication that arises from this is the definition of the yield
stress. For sufficiently low shear rates, foam will spontaneously
release stress, usually as part of coarsening process as bubble
sizes change. This complication is minimized in the bubble raft
given that no substantial coarsening was observed in the absence
of applied shear. In either case, a useful operational definition
of the yield stress is the zero shear-rate limit of the stress.
For the bubble raft of interest here, the average stress as a
function of shear rate is well described by a Herschel-Bulkley
model ($\sigma(\dot{\gamma}) = \tau_o + \mu \dot{\gamma}^n$)
\cite{BAH77}.  From these results, one can determine a yield
stress: $\tau_o = 0.8 \pm 0.1\ {\rm mN/m}$ \cite{PD03,LCD04}. For
the particular shear rate of interest here, this is different from
the ``critical'' stress at which the system begins to undergo
``steady'' flow.

\section{Experimental Results}
\subsection{Elastic Regime} \label{elasticsec}

\begin{figure}
\includegraphics[width=8cm]{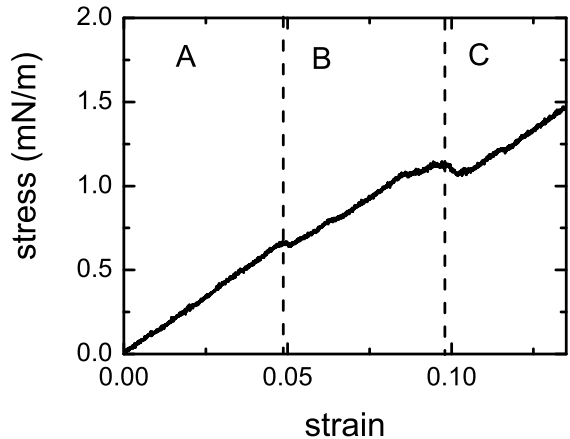}
\caption{\label{initstress} Stress on the inner cylinder versus
strain for the initial period of shear. The curve is divided into
three regions labelled A, B, and C. Region A is the only period
for which no T1 events are observed.}
\end{figure}

The initial stress response of the system is given in
Fig.~\ref{initstress}. There are a number of interesting features
of this regime. First, there are three separate regions of the
initial response, which is essentially set by the slope of the
stress versus strain curve. These regions are indicated by the
vertical dashed lines and are separated by isolated stress-drops.

The first region is the linear, elastic response of the material.
During this period no T1 events are observed. The second two
regions represent plastic deformations in the sense that T1 events
occur. These events are too small to produce stress drops. But,
they modify the slope of the stress-strain curve and generate
irreversible deformation. Hence, the identification of these
regions with plastic deformations. The onset of plastic response
is another useful definition of the yield stress. However, there
is always ambiguity associated with the definition of the onset of
T1 events due to the possibility of T1 events that are the result
of coarsening and not shear. As discussed for the bubble raft,
this difficulty is minimized as no coarsening was observed.
However, for the measurements in Fig.~\ref{initstress}, only a
fraction of the trough was in view. This limits the degree to
which the yield stress can be measured by this method. However, it
is useful to note that the onset of T1 events for the single set
of data studied here is consistent with the value of yield stress
as determined by fits to the behavior of stress as a function of
rate of strain.

The fact that any deviation of the stress-strain curve from linear
behavior is small allows for the definition of an effective shear
modulus of the bubble raft, $G$, during periods of stress
increase. For the initial region in Fig.~\ref{initstress}, $G$ is
the elastic shear modulus. The calculation of $G$ assumes that the
stress is proportional to the strain. The boundary conditions
consist of a fixed rotation rate at the outer boundary and an
inner boundary that is free to rotate, but supported by a torsion
wire. Because of the symmetry of the Couette geometry, the
azimuthal velocity, $v(r)$ is only a function of the radial
position $r$. This is a standard problem; however, given the
slightly unorthodox boundary conditions of this experiment, the
solution is repeated here. The relevant constitutive equation is
\begin{equation}
\label{elastic}
\sigma(r) = G \gamma(r).
\end{equation}
Here $\gamma(r)$ is the shear strain, and $\sigma(r)$ is the
resulting shear stress. In the cylindrical geometry, $\gamma(r) =
r\frac{d\theta(r)}{dr}$, where $\theta(r)$ is the angular
displacement of the bubble raft. For a material confined between
two cylinders, the shear stress is given by $\sigma (r) = T/(2\pi
r^2)$. This follows directly from balancing torques on each
material element. If the bubble raft was a perfectly rigid solid,
one would simply have $v(r) = \Omega r$. This is due to the fact
that the inner boundary is supported by a torsion wire and rotates
as it measures the torque on the inner cylinder. However, for a
finite value of $G$, plugging into Eq.~\ref{elastic}, we get
\begin{equation}
\frac{T}{2\pi r^2} = G\left( r\frac{d\theta(r)}{dr} \right).
\end{equation}
Integrating this equation, and using the fact that the bubble raft
does not slip at either boundary, gives for $G$,
\begin{equation}
\label{Gexp}
G = \frac{\omega}{\Omega - \omega}\left(
\frac{\kappa}{4\pi}\right) \left(\frac{1}{r_i^2} - \frac{1}{R^2}
\right),
\end{equation}
and for $v$,
\begin{equation}
\label{velastic}
 v(r) = \Omega r - \left[
\frac{\kappa\omega r}{4\pi G} \left(\frac{1}{r^2} - \frac{1}{R^2}
\right) \right],
\end{equation}
This equation for $v$ can be rewritten by plugging in for $G$,
\begin{equation}
vr) = \Omega r + \left[ \frac{(\Omega - \omega)r_i^2}{R^2 - r_i^2}
\left(r - \frac{R^2}{r} \right) \right].
\end{equation}
The second piece in the expressions for $v(r)$ is due to the
elastic nature of the bubble raft and the motion of the inner
cylinder. When $\kappa/G$ is small, the system behaves as a rigid
body ($\omega = \Omega$), as expected (large $G$ limit).

\begin{figure}
\includegraphics[width=8cm]{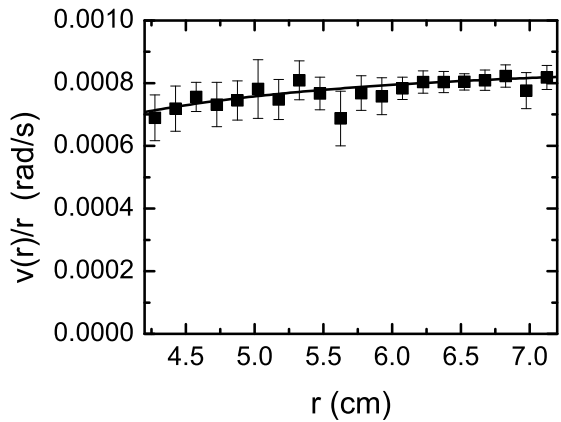}
\caption{\label{elflow} A plot of $v(r)/r$ versus radial position
for the bubble motion during interval A in Fig.~\ref{initstress}.
The solid squares are data averaged over all bubbles at a given
radial position. The solid line is a fit to Eq.~\ref{velastic}.}
\end{figure}

Using the above results, one can find $G$ from measurements of the
average velocity using Eq.~\ref{velastic} or from $\omega$ using
Eq.~\ref{Gexp}. For example, the results for $v(r)$ in region (A)
of Fig.~\ref{initstress} are given in Fig.~\ref{elflow}. The solid
line is a fit to Eq.~\ref{velastic}, with $v(r)/r = 8.2 \times
10^{-4}\ {\rm rad/s} - 0.003\ {\rm rad/cm^2s}(1/r^2 - 0.0193\ {\rm
cm^{-2}})$. The T1 events result in a reduction in the effective
elastic modulus of the bubble raft. The calculated values of $G$
for the three regions are: (A) $G = 11.2 \pm 0.1\ {\rm mN/m}$; (B)
$G = 5.4 \pm 0.1\ {\rm mN/m}$; and (C) $G = 8.9 \pm 0.1\ {\rm
mN/m}$.

\begin{figure}
\includegraphics[width=8cm]{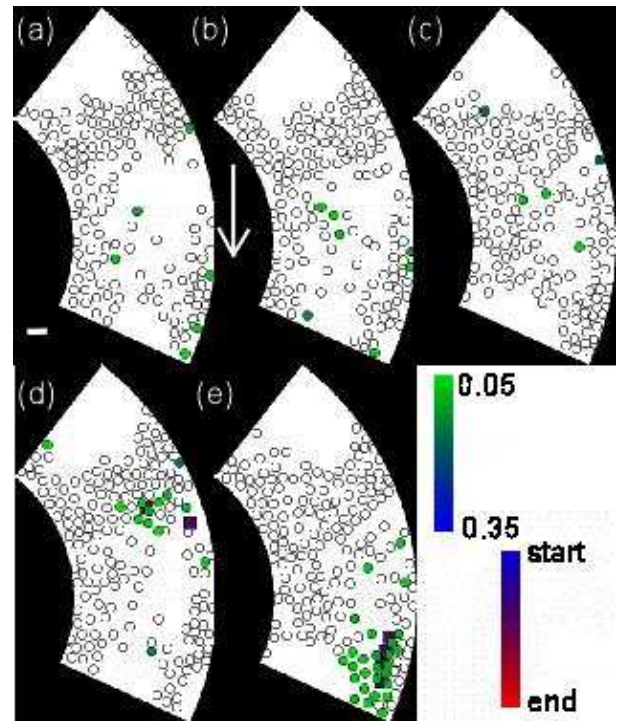}
\caption{\label{t1elastic} Five images representing typical bubble
deviations from elastic flow during the initial stress rise.
Images (a) - (c) are taken from the corresponding three strain
intervals in Fig.~\ref{initstress} and show typical strain
intervals in which no T1 events occur. Image (d) shows a typical
localized T1 event from region (B) in Fig.~\ref{initstress}. Image
(e) shows an event composed of multiple T1 events in which two
rows of bubbles slip pass each other. This event is taken from
region (C) in Fig.~\ref{initstress}. The circles indicate the
location of a subset of bubbles that have been tracked. The sizes
of all the circles are the same, independent of actual bubble
size, for clarity. They are color coded based on the deviation
from elastic displacements, as defined in the text. White
represents deviations less then $0.05\ {\rm cm}$. The color bar
gives the scale for deviations greater than $0.05\ {\rm cm}$. The
squares represent the location of T1 events, where the color
equals the time relative to the start of the interval. The scale
bar in image (a) is 0.5~mm.}
\end{figure}

We use the elastic regime to provide a characterization of the
local deviation from elastic flow. First, we take the fit of
$v(r)$ in region (A) as the definition of ``ideal'' elastic
motion. Knowing this velocity curve, we can compute the expected
displacement of a bubble during a strain interval. From this, we
define $\Delta$ to be a measure of the deviation from elastic
behavior: $\Delta = \sqrt{(x - x_e)^2 + (y - y_e)^2}$, where $x$
and $y$ are the actual displacements of the bubble and $x_e$ and
$y_e$ are the expected displacements if the motion was ``ideal''
elastic behavior. As can be seen from Fig.~\ref{elflow}, even in
the ``pure elastic'' regime there is a significant non-zero
variation to the bubble motions. (The error bars represent the
standard error based on the standard deviation of measured
velocities in each radial bin.) The variation in bubble velocity
is due to a combination of effects, including the obvious fact
that one expects a distribution of displacements due to the finite
size of the bubbles. Therefore, the variation in displacements
from the ideal elastic behavior in region (A) is used to set a
minimum threshold value for $\Delta$ of $0.05\ {\rm cm}$. Bubbles
with a value of $\Delta$ below this threshold are considered to
have undergone ``elastic'' motion. Even with this cutoff, there
are a small number of bubbles in the tails of the distribution
that are classified as deviating from elastic behavior even in
region A. This is illustrated in Fig.~\ref{t1elastic}a-c. Each of
Fig.~\ref{t1elastic}a-c, represent a period of strain of 0.064 in
which no T1 events occur. The periods were selected from the
corresponding region (A - C) of Fig.~\ref{initstress}. The circles
indicate the location of tracked bubbles (so only a fraction of
the total bubbles are shown). The color of the bubbles indicates
the deviation from elastic behavior, with white bubbles having a
value of $\Delta < 0.05\ {\rm cm}$. The color code is indicated in
the figure.

Figure~\ref{t1elastic}d and e illustrate two classes of T1 events
that do not result in a stress drop. Fig.~\ref{t1elastic}d is from
region (B) of Fig.~\ref{initstress}. This illustrates an isolated
pair of T1 events that have an associated region in which the
bubbles deviate from elastic behavior. Fig.~\ref{t1elastic}e is
from region (C) of Fig.~\ref{initstress}. This illustrates the
slippage of two, short rows of bubbles due to simultaneous T1
events. Again, there is a relatively localized region of deviation
from elastic behavior associated with these T1 events.

\subsection{Plastic Flow Regime} \label{plasticsec}

In the plastic flow regime, there are two main questions regarding
the T1 events. First, what is the correlation between T1 events
and stress? Second, what is the correlation between T1 events and
bubble motion, as measured by either the average velocity or the
deviation from elastic behavior?

\begin{figure}
\includegraphics[width=8cm]{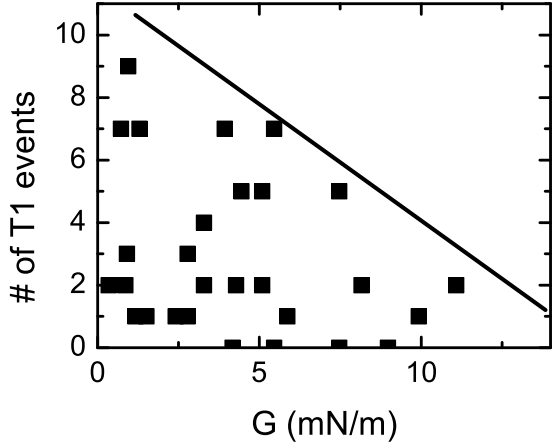}
\caption{\label{GvT1}
Scatter plot of the number of T1 events during a period of stress
increase versus the effective elastic modulus $G$ for the same
period of strain. The solid line is a guide to the eye.}
\end{figure}

Regarding the correlation between T1 events and stress, it is
interesting to consider the periods of stress increase. As with
region B and C in Fig.~\ref{initstress}, there are often T1 events
during these periods of stress increase. Therefore, in general,
these are periods of plastic deformation, though preliminary
observations suggest that occasional increases exist during which
no T1 events occur. One way to characterize stress increases is to
use Eq.~\ref{Gexp} to calculate an effective shear modulus, $G$,
for each separate period of stress increase. This can than be
correlated with the number of T1 events that occur in that period.
A preliminary measurement of this is shown in Fig.~\ref{GvT1}.
This result is preliminary because only a fraction of the sample
was viewed. Therefore, the results for the number of T1 events
represent a lower bound. However, it is interesting that the data
all falls below the straight line, suggesting a correlation
between $G$ and the number of T1 events, as expected.

\begin{figure}
\includegraphics[width=8cm]{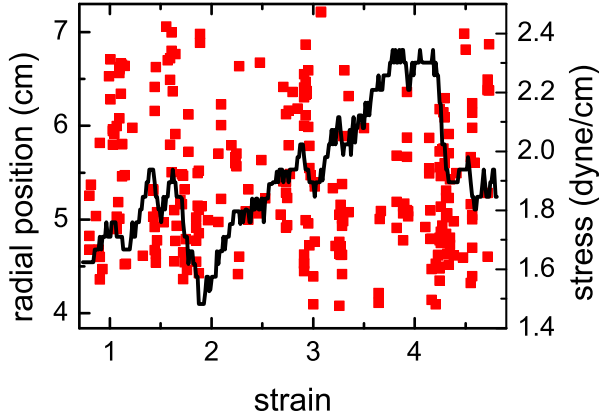}
\caption{\label{t1pos} The individual points are the radial
position of T1 events as a function of strain. The solid curve is
the stress as a function of strain (measured using images of the
inner cylinder).}
\end{figure}

It is natural to expect that one necessary condition for a T1
event to occur is that the local stress exceeds some critical
value. This would suggest a correlation between the stress and the
location of the T1 events. Figure~\ref{t1pos} illustrates that no
correlation exists between the stress on the {\it inner cylinder}
and the radial positions of T1 events. One would expect such a
correlation if the stress field was given by the continuum limit,
which in a Couette geometry is $\sigma(r) = (\sigma(r_i)
r_i^2)/r^2$, and the critical stress for a T1 event was spatially
uniform. Under these conditions, for each $\sigma(r_i)$, there is
a maximum $r$ at which T1 events can occur. This is set by the
critical stress required for the generation of a T1 event. The
lack of a correlation suggests that at least one, if not both, of
these assumptions is false. In fact, work in other systems
suggests that both assumptions are false. Given the direct
measurement of stress chains in granular matter \cite{HBV99}, it
is reasonable to expect such chains in the bubble raft. This would
represent a breakdown of the continuum assumption for the stress
distribution. Also, simulations of amorphous metal have shown the
existence of localized, high stress regions (referred to as
$\tau$-defects) that are the sources of local flow \cite{SVE83}.
In other contexts, models and simulations have suggested the
existence of ``weak'' zones in complex fluids
\cite{LRMGW96,SLHC97,LL97} that are the source of viscous-like
behavior. A more detailed study of these issues will required
improved images, but the current work is very suggestive.

\begin{figure}
\includegraphics[width=8cm]{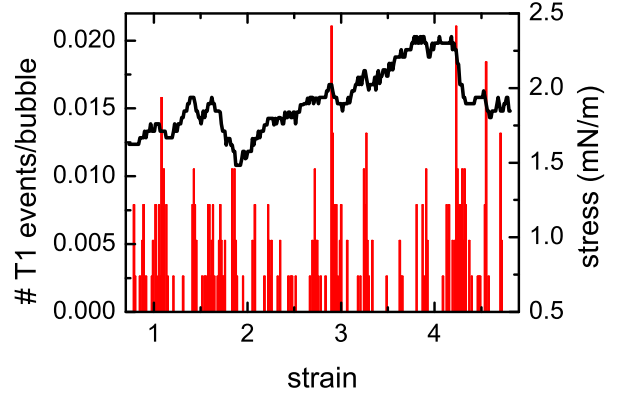}
\caption{\label{t1rate} The solid line is the same stress versus
strain curve as shown in Fig.~\ref{t1pos}. The bars summarize the
data in Fig.~\ref{t1pos} by plotting only the number of T1
events/bubble in a strain interval of $0.013$.}
\end{figure}

To summarize the average properties of T1 events as a function of
strain, Fig.~\ref{t1rate} plots the number of T1 events per bubble
versus strain. Again, this is shown simultaneously with the stress
versus strain curve to illustrate the general correlation between
the size of the stress drops and the total number of T1 events.
One observes that most stress drops consist of a cascade of events
throughout the stress drop. As with the stress increases, detailed
correlations between the size of a stress drop and the number of
T1 events will require images of the entire sample. However, one
can compute $R_{T1}$. For this shear rate, $R_{T1} = 0.18 \pm
0.01$, in reasonable agreement with both the bubble model and
Langmuir monolayer foam.

The next question is the connection between velocity profiles and
T1 events. Based on the results of Ref~\cite{LCD04}, it is known
that there exists a shear discontinuity at $r_c = 6.7\ {\rm cm}$
for the system reported on here. Therefore, there is no
expectation of strong localization of the T1 events are reported
in Ref.~\cite{KD03} because there is no shear localization.
However, one might expect a connection between the radial
distribution of T1 events and the shear discontinuity.

The shear discontinuity divides the system into two regimes. Below
$r_c$, the average velocity is consistent with that of a power-law
fluid. Above $r_c$, the systems acts like an elastic solid.
Figure~\ref{t1vel} illustrates the connection between the average
velocity profile and the spatial distribution of T1 events. The
vertical line indicates the spatial location of the shear
discontinuity \cite{LCD04}. The basic shape of the distribution is
similar to that found in the simulations reported in
Ref.~\cite{KD03}. There is a ``peak'' at smaller radii, with the
distribution tailing off as one goes to larger radii. The main
difference is the location of the cutoff in the T1 distribution.
As reported in Ref.~\cite{KD03}, the cutoff in velocity and T1
events is at essentially the same radius. In contrast, for the
system reported on here, there is no obvious signature in the
distribution of T1 events at the shear discontinuity (see
Fig.~\ref{t1vel}). This may be due to the fact that even though
the shear-rate is zero, the bubbles are still moving near the
outer wall, and differences in bubble size may lead to T1 events.
Also, it may be an artifact of how close the shear discontinuity
is to the outer wall. Future work in larger systems is needed to
better understand this issue.

In order to better understand the detailed connection between T1
events and stress drops, two short periods of strain are
highlighted, as indicated in Fig.~\ref{t1highlight}. These are
segments of the data presented in Fig.~\ref{t1pos}. The period of
strain illustrate in Fig.~\ref{t1highlight}a was selected to
highlight the nature of potential correlations between the stress
behavior and the number of T1 events. First, the initial elastic
rise A is included for comparison with the stress increase in the
interval labelled C. During the initial rise, there is only one
observed T1 event, and the effective elastic modulus is $G = 5.6
\pm 0.1\ {\rm mN/m}$. In contrast, during region C, there are 9
observed T1 events, and the effective elastic modulus is $G = 2.1
\pm 0.1\ {\rm mN/m}$. These results reinforce the general
connection between number of T1 events and effective elastic
modulus discussed with respect to Fig.~\ref{GvT1}. In contrast,
the regions labelled B - E all have roughly the same number of T1
events. Yet, region B and E are stress drops. Region C is a period
of stress increase, and region D is a slight decrease in stress.
One difficulty in drawing definitive conclusions from this data is
the fact that only part of the system is being viewed. However,
this strongly suggests that the additional bubble motions, not
just the T1 events, play an important role in determining the
overall stress evolution.

The interval illustrated by Fig.~\ref{t1highlight}b was selected
to make connections with the velocity profiles reported in
Ref.~\cite{LCD04} in an attempt to better understand the shear
discontinuity that occurs at $r_c$. Here, the average bubble
displacements are measured, but these are directly related to
average velocities. This sequence is particularly interesting
because there are three stress drops that occur at different
average stress values, and the drop at the lowest average stress
(region E) exhibits the larger value of $r_c$ \cite{LCD04}. (We
are considering the behavior in region B and C as two separate
stress drops because of the short plateau between them. However,
this points out the issue regarding the definition of ``events''
as discussed earlier in the context of Ref.~\cite{ML04}.)

The spatial distribution of T1 events and bubble deviations from
elastic behavior are given in Fig.~\ref{t1flow} using the same
criteria as described for Fig.~\ref{t1elastic} in
Sec.~\ref{plasticsec}. White bubbles represent essentially elastic
behavior, and the color of the other bubbles is the degree to
which their motion deviates from elastic. The letters in
Fig.~\ref{t1highlight}b correspond to the labelling of
Fig.~\ref{t1flow}.  One observes very similar distributions of T1
events for all three stress drops (see Fig.~\ref{t1flow}b,c, and
e). If one looks carefully, the distinguishing factor appears to
be the number of bubbles deviating from elastic behavior at any
given radius. This is made clearer by considering the average
bubble displacement as a function of radial position, as
illustrated in Fig.~\ref{bubdispl}.

Figure~\ref{bubdispl} is a plot of $\Delta \theta/ \Omega \Delta
t$ versus radial position. The values of $\Delta \theta$ are
computed by dividing the system into 20 equally spaced radial bins
and averaging the angular displacements over all bubbles in a
given bin over the time interval of interest. The time intervals
are selected so that they match the strain intervals indicated in
Fig.~\ref{t1highlight}b. For comparison, the displacements during
the essentially flat regions in stress are given as open symbols,
and the displacements during the stress drops are given as closed
symbols. The angular displacement is normalized by $\Omega\Delta
t$. The solid line is the ``ideal'' elastic behavior given by the
fit to the data in Fig.~\ref{elflow}. One can see that for
interval E (solid triangles), the deviation from the expected
elastic displacement occurs at the largest value of $r_c$. This is
consistent with the results reported in Ref.~\cite{LCD04} for the
velocity profiles. What is new in these results is the ability to
correlate the location of T1 events during a stress drop and the
location of the deviation from elastic behavior.

For event E, there are two clear deviations from elastic behavior,
as shown in Fig.~\ref{bubdispl}: (a) at $r = 6.48\ {\rm cm}$ there
is a positive deviation; and (b) at $r = 6.16\ {\rm cm}$ there is
a negative deviation. During the drop, the maximal radial position
of a T1 event is $r = 6.24\ {\rm cm}$. The existence of positive
and negative deviations is consistent with the bubbles associated
with a T1 event moving both forward and backward relative to the
average flow. The average displacements during B and C are
essentially identical. However, for C more then B there is some
evidence for a positive and negative deviation at $r = 5.84\ {\rm
cm}$ and at $r = 5.44\ {\rm cm}$, respectively. For these drops,
the maximal radial position of a T1 event is $r = 6.29\ {\rm cm}$.
Comparing these numbers strongly suggests that the location of T1
events is not the main contribution to the determination of the
deviation from elastic behavior, and hence, the determination of
$r_c$. Instead, it is the detailed motion of the surrounding
bubbles. Interestingly, the greatest difference between the two
events in terms of T1 position is in the precursor to the drops;
yet the precursors have very similar angular displacements (open
symbols in Fig.~\ref{bubdispl}). During the interval labelled D in
Fig.~\ref{t1highlight}b, one observes T1 events as far out as $r =
6.98\ {\rm cm}$. In contrast, during the interval labelled A in
Fig.~\ref{t1highlight}b, one only observes T1 events as far out as
$r = 5.87\ {\rm cm}$. Presumably these events play an important
role in establishing the local stress fields that govern the
bubble motions during the subsequent stress drop.

\section{Discussion} \label{summary}

Even though various aspects of the work presented here are
preliminary in the sense that only a portion of the entire raft
was studied, a number of questions regarding the role of T1 events
in the macroscopic response of a bubble raft to flow have been
addressed. First, the contribution of T1 events to the effective
shear modulus was considered. T1 events during periods of stress
increase effectively lower the shear modulus of the bubble raft.
Similarly, during stress drops, there is a correlation between the
size of the drop and the total number of T1 events. Future work is
required to establish a more detailed correlation between the
number of T1 events and the effective shear modulus and size of
stress drops.

\begin{figure}
\includegraphics[width=8cm]{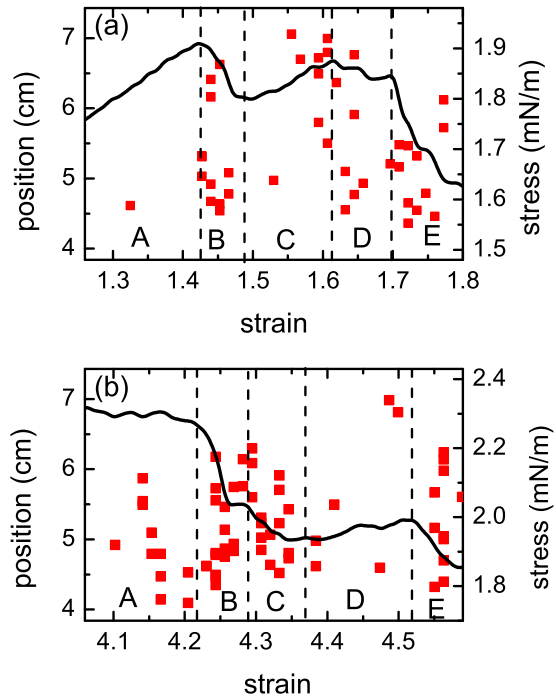}
\caption{\label{t1highlight} This is two strain intervals from
Fig.~\ref{t1pos}, showing both the location of T1 events (solid
squares) and the stress (solid line, as measured by the magnetic
method) as a function of strain. Each interval is further divided
into smaller strain intervals by the vertical lines. The labels in
(b) correspond to the images in Fig.~\ref{t1flow}.}
\end{figure}

\begin{figure}
\includegraphics[width=8cm]{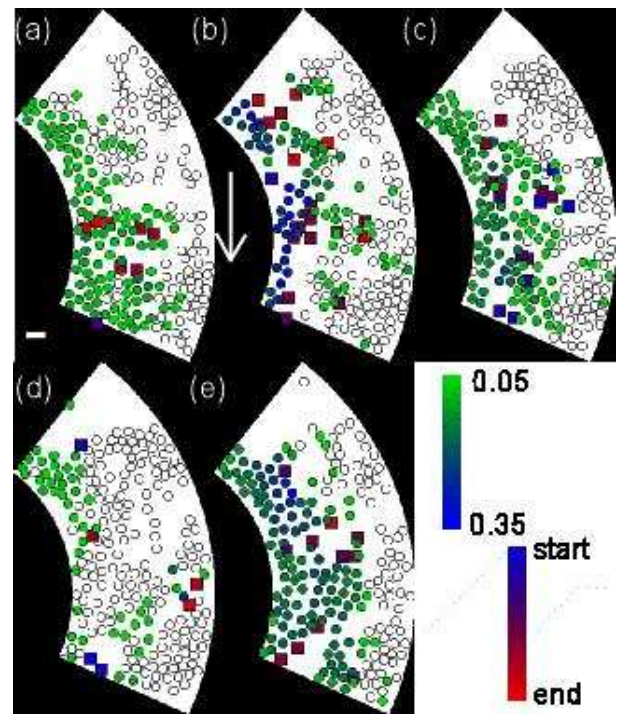}
\caption{\label{t1flow} Five images representing typical bubble
deviations from elastic flow during the corresponding strain
intervals as indicated in Fig.~\ref{t1highlight}(b). The circles
indicate the location of a subset of bubbles that have been
tracked. The size of the circle is the same for all bubbles for
clarity. The circles are color coded based on the deviation from
elastic displacements, as defined in the text. White represents
deviations less then $0.05\ {\rm cm}$. The color bar gives the
scale for deviations greater than $0.05\ {\rm cm}$. The squares
represent the location of T1 events, where the color equals the
time relative to the start of the interval. The scale bar in image
(a) is 0.5~mm.}
\end{figure}

Correlations between positions of T1 events, average stress,
individual bubble displacements, and average angular displacements
of bubbles were considered. A general picture that emerges from
these measurements is the importance of understanding the local
stress field and the local geometry of bubbles. For example, an
investigation of individual bubble motions before and during a
stress drop (see discussion of Figs.~\ref{t1highlight} and
\ref{t1flow}) shows that the radial distribution of T1 events can
not be understood in terms of a simple continuum model and single
stress threshold. The T1 events in the strain period immediately
prior to a stress drop play an important role in establishing the
local stress field and geometric relations between bubbles that
sets the subsequent motions. For example, the two different stress
drops highlighted in Fig.~\ref{t1highlight} exhibited similar
distributions of T1 events, but the deviations from elastic
behavior and the average displacements were very different. The
main differences between the events was in the distribution of
precursor T1 events, not in the average bubble motions.

\begin{figure}
\includegraphics[width=8cm]{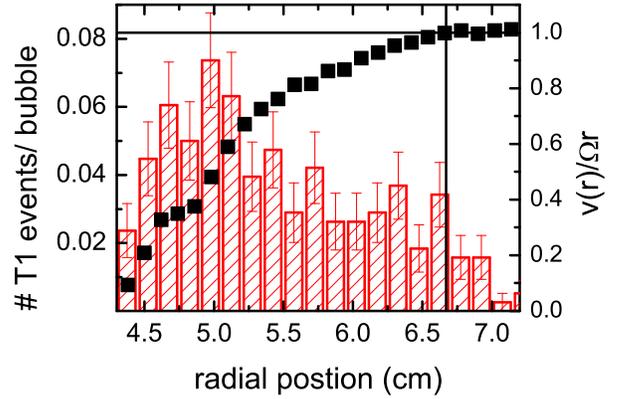}
\caption{\label{t1vel} The squares are the average velocity of the
bubbles normalized by $\Omega r$ as a function of radial position.
The bars give the total number of T1 events/bubble as a function
of radial position. The solid lines are guides to the eye. The
horizontal line is $v(r)/\Omega r = 1$, which corresponds to the
motion of a rigid body. The vertical line is the location of the
shear discontinuity, as reported in Ref.~\cite{LCD04}.}
\end{figure}

\begin{figure}
\includegraphics[width=8cm]{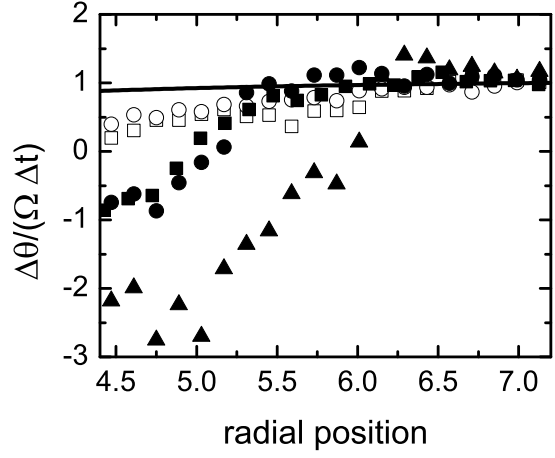}
\caption{\label{bubdispl} The average angular displacement of the
bubbles normalized by $\Omega \Delta t$ as a function of radial
position. The different symbols are for the different strain
intervals in Fig.~\ref{t1highlight}b: (A) open circles; (B) solid
squares; (C) solid circles; (D) open squares; and (E) solid
triangles. Here $\Delta t$ is the time interval for each strain
interval. The solid line is the fit to elastic behavior from
Fig.~\ref{elflow}.}
\end{figure}

The connection between T1 events and the position of the shear
discontinuity was also considered. Both in terms of the average
properties (see Fig.~\ref{t1vel}) and the short time motions (see
Fig.~\ref{bubdispl}). There is no clear evidence for a connection
between the positions of T1 events and the shear discontinuity,
but larger system sizes need to be studied. However, there may be
an indirect connection through the stress relaxation and
subsequent motion of surrounding bubbles. Indirectly, these
measurements have some potential implications for the simulations
of the modified vertex model \cite{KD03}. These simulations
illustrate that a localization of T1 events can lead to a shear
localization. This system does not exhibit localization of either
the T1 events or the shear. This indirectly supports the
connection between T1 event localization and shear localization.
What remains an important questions is why would T1 events
localize in one case and not the other? An obvious difference
between the T1 events in the bubble raft and in the simulation is
the duration of the T1 events. In the model, the T1 events all
occur on a very short time scale, by construction. For the bubble
raft, there is a distribution of times for the duration of T1
events. Some events occur very slowly (over 10 - 20 seconds). It
is these difference in duration that may modify the impact on the
local stress. Again, this point to the importance of understanding
the local stress fields generated by the T1 events, and not just
the distribution of the events themselves. Furthermore, as part of
the future work that focuses on local stress fields, it will be
important to correlate the changes in local stress with the
duration of the T1 events.

The measurements reported here focused on bubble displacements and
T1 events. Where possible, comparisons with the bubble model show
quantitative agreement, such as for $R_{T1}$. This adds support to
previous results with the bubble raft that were also in general
agreement with the bubble model \cite{LTD02,PD03}. Having strong
agreement between the experiments and a theoretical model is
useful for the next stage of experimental studies. Essentially all
of the results point to the need for measurements of the local
stress field. Future experimental work is planned that will use
the bubble distortion as a direct measure of local stress, as has
been done with other foam systems \cite{RK00,BB02b,AJGG03,JG03}.
Close contact with simulations that focus on the stress released
by T1 events and STZ's, as well as experimental studies of
granular material, will be important for understanding this future
work.

\begin{acknowledgments}

This work was supported by the Department of Energy grant
DE-FG02-03ED46071, the Research Corporation, and the Alfred P.
Sloan Foundation. I thank John Lauridsen for use of his video data
of bubble rafts that was taken while an undergraduate at the
University of California, Irvine. I thank Craig Maloney, Corey
O`Hern, Michael Falk, and Georges Debr\'{e}geas for fruitful
discussions.

\end{acknowledgments}

%\bibliography{d04}

\begin{thebibliography}{50}
\bibitem[{tes()}]{test}
\bibinfo{note}{The detection of T1 events by hand was tested using data
  generated based on the bubble model. The number of T1 events detected by hand
  and with the computer agreed well within expected statistical errors.}
\end{thebibliography}

\clearpage

\end{document}